\documentclass[iicol]{sn-jnl}
\usepackage[switch,mathlines]{lineno}
\usepackage{hyperref}
\usepackage{doi}
\usepackage{uri}
\usepackage{amsmath}
\usepackage{amssymb}

\usepackage{float}
\usepackage{graphicx}

\let\oldequation\equation
\let\oldendequation\endequation

\renewenvironment{equation}
  {\linenomathNonumbers\oldequation}
  {\oldendequation\endlinenomath}
\begin{document}

\title[Article Title]{Fast generation of Schrödinger cat states in a Kerr-tunable superconducting resonator}


\author[1,2]{\fnm{X.L.} \sur{He}}
\equalcont{These authors contributed equally to this work.}
\author*[3,4]{\fnm{Yong} \sur{Lu}}\email{kdluyong@outlook.com}
\equalcont{These authors contributed equally to this work.}

\author[1,2]{\fnm{D.Q.} \sur{Bao}}

\author[1,2]{\fnm{Hang} \sur{Xue}}

\author[1,2]{\fnm{W.B.} \sur{Jiang}}

\author[1,2]{\fnm{Z.} \sur{Wang}}
\author[4]{\fnm{A.F.} \sur{Roudsari}}
\author[4]{\fnm{Per} \sur{Delsing}}

\author[5,6]{\fnm{J.S.} \sur{Tsai}}

\author*[1,2]{\fnm{Z.R.} \sur{Lin}}\email{zrlin@mail.sim.ac.cn}

\affil[1]{\orgdiv{National Key Laboratory of Materials for Integrated Circuits}, \orgname{Shanghai Institute of Microsystem and Information Technology, Chinese Academy of Sciences},  \city{Shanghai}, \postcode{200050},  \country{China}}

\affil[2]{\orgname{University of Chinese Academy of Science}, \city{Beijing}, \postcode{100049}, \country{China}}

\affil[3]{\orgdiv{3rd Physikalisches Institut}, \orgname{University of Stuttgart},  \postcode{70569}, \state{Stuttgart}, \country{Germany}}

\affil[4]{\orgdiv{Microtechnology and Nanoscience}, \orgname{Chalmers University of Technology}, \postcode{SE-412 96}, \state{Göteborg}, \country{Sweden}}

\affil[5]{\orgdiv{Graduate School of Science}, \orgname{Tokyo University of Science},  \city{Shinjuku}, \state{Tokyo}, \postcode{162-0825}, \country{Japan}}

\affil[6]{\orgdiv{Center for Quantum Computing}, \orgname{RIKEN},  \city{Wako}, \state{Saitama}, \postcode{351-0198}, \country{Japan}}

\abstract{Schrödinger cat states, quantum superpositions of macroscopically distinct classical states, are an important resource for quantum communication, quantum metrology and quantum computation. Especially, cat states in a phase space protected against phase-flip errors can be used as a logical qubit. However, cat states, normally generated in three-dimensional cavities, are facing the challenges of scalability and controllability. Here, we present a novel strategy to generate and store cat states in a coplanar superconducting circuit by the fast modulation of Kerr nonlinearity. At the Kerr-free work point, our cat states are passively preserved due to the vanishing Kerr effect. We are able to prepare a 2-component cat state in our chip-based device with a fidelity reaching 89.1\% under a 96\ ns gate time. Our scheme shows an excellent route to constructing a chip-based bosonic quantum processor.}




\maketitle

\begin{figure}[h]
\centering
\includegraphics[width=0.45\textwidth]{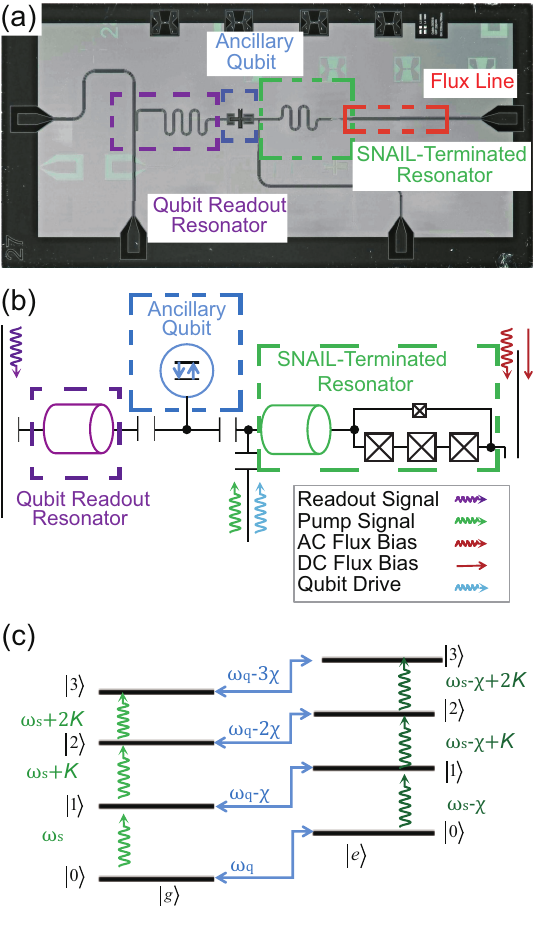}
\caption{Structure of the superconducting circuit. (a) A microscopic photo of the superconducting circuit. An ancillary qubit in the middle is capacitvely coupled to both a readout resonator (left) and a SNAIL-terminated resonator (right). (b) Schematic circuit diagram of the system. (c) Energy structure of the dispersively coupled nonlinear resonator and qubit.  $\lvert g \rangle$ and $\lvert e \rangle $  are the ground and excited state of qubit respectively. $\lvert 0 \rangle,\lvert 1 \rangle,\lvert 2 \rangle,...$represent the energy levels of the resonator. $\chi$ is the dispersive shift. }\label{Fig1}
\end{figure}

Quantum computation has been proven to surpass classical architectures in certain computational tasks \cite{b1}. Quantum information has been encoded and manipulated in diverse systems such as cold atoms \cite{b2}, trapped ions \cite{b48}\cite{b49}, superconducting circuits \cite{b50}. Especially, superconducting circuit is a promising platform which has shown significant progress on the gate-based quantum computers \cite{b1}\cite{b5}. Additional qubit elements are normally required to achieve large-scale error-correctable two-level system based quantum computation \cite{b51}. In contrast, the phase space of a bosonic system inherently provides a larger Hilbert space and thus a larger coding area \cite{b6}\cite{b7}\cite{b8}\cite{b42}. Therefore, encoding quantum information in continuous-variables leads to a significant reduction in hardware overhead on the path towards the fault-tolerance. The nonclassical states with negative Wigner functions \cite{b22}\cite{b23} can be regarded as a quantum computing resource to obtain quantum computational advantage. Recently, nonclassical states including Schrödinger's cat codes\cite{b57} binominal codes \cite{b11}, GKP states \cite{b12}\cite{b13}, and cubic-phase states \cite{b12}\cite{b52}, have been demonstrated in cavities coupled to ancillary qubits. However, the ancillary qubit normally has a fixed Kerr nonlinearity which might be detrimental even for the storage of nonclassical state \cite{b15}. In previous results, non-classical states were mostly observed in 3-dimensional superconducting cavities based on the Kerr nonlinearity \cite{b15}\cite{b18}, and the logic states were implemented by engineering the two-photon losses \cite{b16}.

In this paper, rather than engineering the two-photon dissipation where adiabatic evolution is required to prepare the states, we demonstrate a fast pulsed scheme to generate Schrödinger's cat states in a coplanar superconducting resonator terminated by a Superconducting Nonlinear Asymmetric Inductive eLement (SNAIL) \cite{b20} as shown in Fig. \ref{Fig1}(a). In contrast to standard linear resonators, our nonlinear resonator enables a large tunable range of the Kerr nonlinearity by changing the external magnetic flux through the SNAIL element It allows us to prepare cat states. Moreover, the four-wave mixing term is zero by tuning the flux bias to the Kerr-free point. We can also preserve the prepared cat states, because the Kerr-induced evolution is prevented. Here, the Wigner functions of the generated states are measured through an ancillary transmon qubit to characterize the bosonic quantum states stored in the nonlinear resonator. 

\begin{figure*}
\centering
\includegraphics[width=0.9\textwidth]{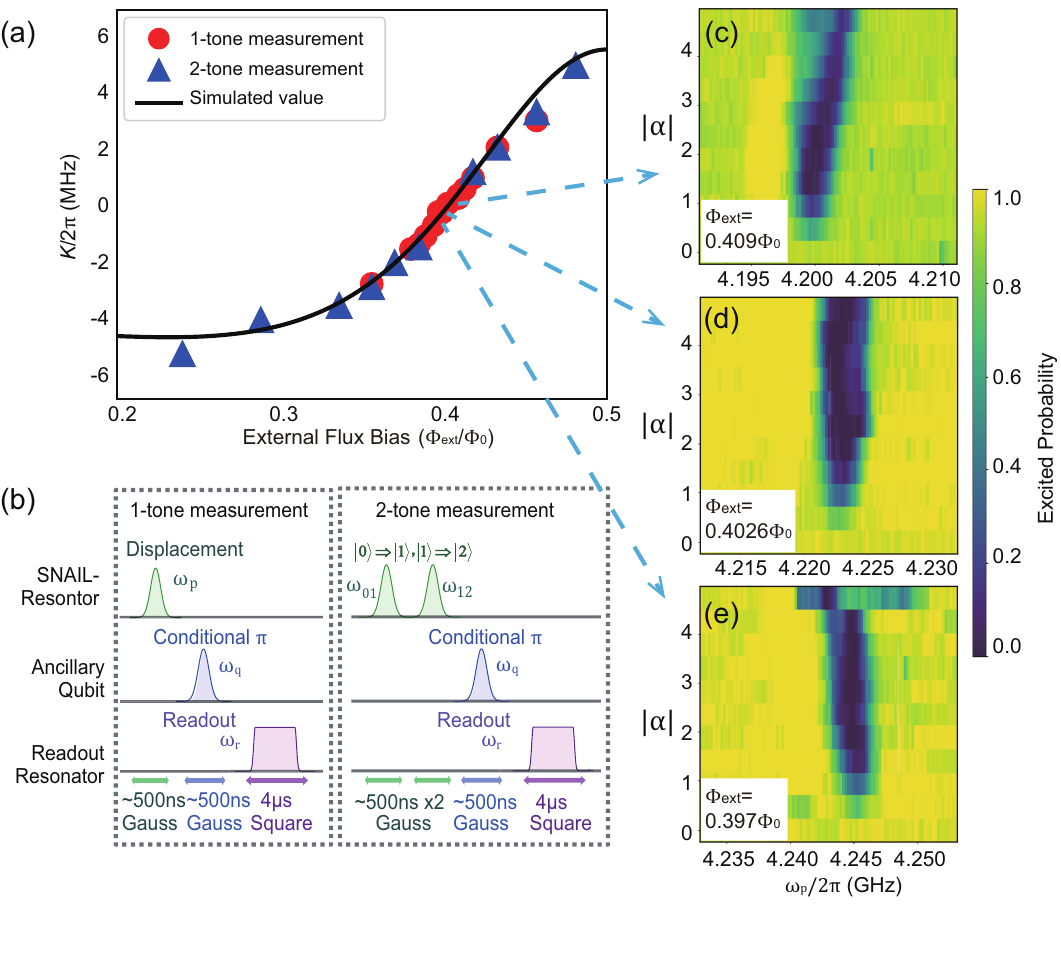}
\caption{Experimental methods to calibrate the tunable nonlinearity of the resonator. (a) Flux bias dependent nonlinearity. The Kerr coefficient $K$ is measured with single-tone and two-tone measurements. (b) Pulse sequences for the single-tone and two-tone nonlinearity measurements. (c)-(e) Results of the single-tone measurement near the Kerr-free point. $\alpha$ is the displacement. $\omega _{\rm p}$ is the frequency of the pump pulse to the SNAIL-terminated resonator.}\label{Fig2}
\end{figure*}
The energy of the SNAIL in our circuit with three big junctions and one smaller junction [Fig. \ref{Fig1}(c)] can be written as \cite{b27}:
\begin{equation}
\begin{split}
{U_{\rm SNAIL}}(\varphi )=- \beta {E_{\rm J}}\cos (\varphi )-3{E_{\rm J}}\cos (\frac{{{\varphi _{\rm ext}}- \varphi }}{3}),\label{eq1}
\end{split}
\end{equation}
where the ratio of the Josephson energies of the small and the big junctions of SNAIL, $\beta\approx0.095$, the Josephson energy $E_{\rm J}/h\approx830\ {\rm GHz}$ \cite{b30} (Details in Methods), $\varphi_{\rm ext}=2\pi\Phi_{\rm ext}/\Phi _{\rm 0}$ is the phase induced by the external magnetic flux and $\varphi$ is the phase difference between two ports of the SNAIL. The Hamiltonian of the SNAIL-terminated resonator is \cite{b8}\cite{b20}\cite{b43}:
\begin{equation}
{H_{\rm SNAIL-Res}} = \hbar {\omega _{\rm s}} a^\dag a + {g_{\rm 3}}{(a + a^\dag )^{\rm 3}} + {g_{\rm 4}}{(a + a^\dag )^{\rm 4}},\label{eq2}
\end{equation}
where $\omega _{\rm s}$ is the resonant frequency of the SNAIL-terminated resonator (the tunable range of $\omega _{\rm s}/2\pi$ is around 4.08-5.00\ GHz in our device). $a$ ($a^\dag$) is the annihilation (creation) operator. $g_{\rm 3}\ (g_{\rm 4})$ is the coupling strength for the three (four)-wave mixing.

\begin{figure*}
\centering
\includegraphics[width=0.9\textwidth]{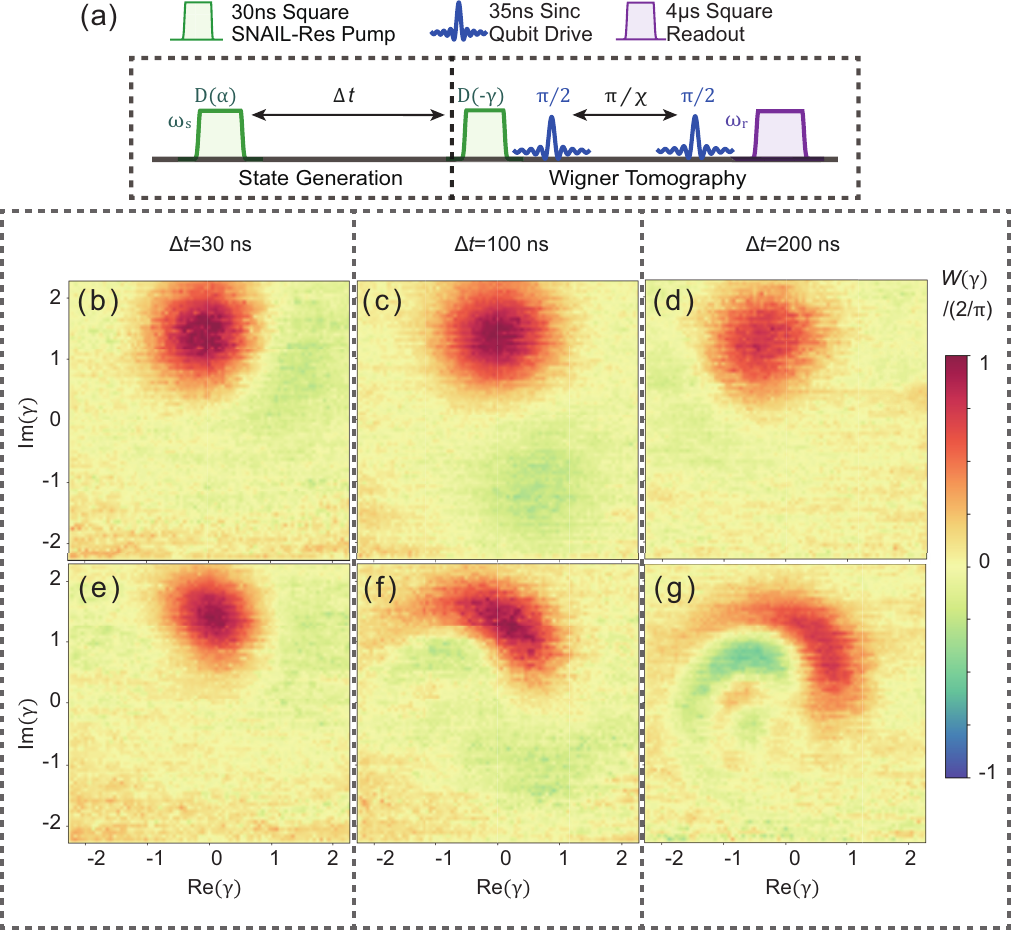}
\caption{Time evolution of a coherent state in the SNAIL-terminated resonator. (a) The pulse sequence where the pulse width of displacement is 30ns, and the time between two displacement pulses is $\Delta {{t}}$. The experimental Wigner tomography shows the evolution progress of the coherent states at different time durations as shown in (b)-(d) at the Kerr-free point and (e)-(g) with $K/2\pi \approx 0.5 {\rm MHz}$ where the Kerr effect clearly distorts the state.}\label{Fig3}
\end{figure*}

Including the coupled ancillary transmon qubit, the total effective Hamiltonian of the system in the dispersive regime is given by \cite{b25}:
\begin{equation}
\begin{split}
\frac{{{H_{\rm eff}}}}{\hbar } \approx \omega_{\rm s}{a^\dag }a + K{a^\dag }^{\rm 2}{a^{\rm 2}}+\frac{{{\omega _{\rm q}}}}{2}{b^\dag }b \\\
- \frac{\chi }{2}{a^\dag }a{b^\dag }b - \frac{{{K_{\rm q}}}}{2}{b^\dag }^{\rm 2}{b^{\rm 2}},\label{eq3}
\end{split}
\end{equation}
where $\omega _{\rm q}$ is the frequency of the ancillary qubit (around 5.09-5.19\ GHz). $b$ ($b^\dag$) is the lowering (raising) operator for the ancillary qubit. $K$ is the Kerr nonlinearity of the resonator, defined as the frequency shift per photon, $K = {K_{\rm s}} + {K_{\rm qs}}$ with the self-Kerr term ${K_{\rm s}} = 12({g_{\rm 4}} - 5g_{\rm 3}^{\rm 2}/{\omega _{\rm s}})$ from the SNAIL element and the cross-Kerr term ${K_{\rm qs}} ={\chi ^{\rm 2}}/4{K_{\rm q}}$ from the qubit with the dispersive shift $\chi/2\pi \approx 3.5-18\ {\rm MHz}$ depending on the flux bias [especially $\chi/2\pi  \approx 4.35\  {\rm MHz}$ when the external flux $\Phi_{\rm ext}  = 0.4026\ {\Phi _{\rm 0}}$(see Methods)]. The qubit anharmonicity is $K_{\rm q}/2\pi \approx - 420\ {\rm MHz}$. The value of $K_{\rm s}/2\pi$ can be tuned from negative to positive with a range up to a few MHz [Fig. \ref{Fig2}(a)], whereas the value of $K_{\rm qs}/2\pi$ is always negative on the order of kHz. Therefore, it is possible to cancel the cross-Kerr term from the qubit to obtain $K=0$ by tuning $K_{\rm s}$ with the magnetic flux through the SNAIL \cite{b39}.

\begin{table}[!ht]
\caption{Parameters of the circuit. ($\omega_{\rm q0}$, $\omega_{\rm s0}$, and $\chi_{\rm 0}$ are the values at Kerr-free working point.)}
\begin{tabular}{|c|c|c|c|}\hline
$\omega_{\rm q}/2\pi$ & 5.09 - 5.19\ GHz & $\omega_{\rm s}/2\pi$ & 4.08 - 5.00\ GHz \\ \hline
$K_{\rm q}/2\pi$ & -420\ MHz & $K_{\rm s}/2\pi$ & (-5) - 6\ MHz \\ \hline
$E_{\rm J}/h$ &  830\ GHz & $\beta$ & 0.095 \\ \hline
$\chi/2\pi$ & 3.5 - 18\ MHz & $\omega_{\rm q0}/2\pi$ & 5.095\ GHz \\ \hline
$\omega_{\rm s0}/2\pi$ & 4.223\ GHz & $\chi_{\rm 0}/2\pi$ & 4.35\ MHz  \\ \hline
\end{tabular}
\end{table}

Firstly, we calibrate the values of the Kerr coefficient $K$ [Fig. \ref{Fig2}(a)] precisely with two approaches, namely single-tone and two-tone measurements [Fig. \ref{Fig2}(b)]. For the single-tone measurement, we sweep the frequency of a displacement pulse $D(\alpha )$ followed by a conditional qubit $\pi$-pulse, where the qubit is excited only if the SNAIL-terminated resonator is empty ($\alpha$ is the displacement with photon number $N=\alpha^2$). Therefore, we can observe the resonator frequency shift with the photon number inside as shown in Fig. \ref{Fig2}(c-e). We can extract the Kerr coefficient $K$ by linearly fitting the relationship between the frequency shift and the photon number (see Methods). This method is valid only for a small $K$ so that the total frequency shift is not larger than the pulse linewidth. For a larger $K$, we switch to perform a two-tone measurement. In this measurement, we regard the nonlinear resonator as a multi-level system with an anharmonicity (similar to a qubit), where we perform Rabi oscillations on the lowest three levels by applying two pulses on the transition $\lvert 0 \rangle  \Rightarrow \lvert 1 \rangle$ and $\lvert 1 \rangle  \Rightarrow \lvert 2 \rangle$, respectively. Thus, the anharmonicity, corresponding to the value of $K$, can be obtained as soon as the resonant frequencies are found (see Methods).

As shown in Fig. \ref{Fig2}(a), the dynamic range of the Kerr coefficient $K/2\pi$ is approximately from -5 MHz to 6 MHz which is close to the theoretically simulated result \cite{b27}. Particularly, with flux bias $\Phi_{\rm ext}  = 0.4026\ {\Phi _{\rm 0}}$, we find a working point where $K$ is small, $\lvert K/2\pi\rvert < 70\ {\rm kHz}$ from single-tone measurement. The accuracy is limited by the spectroscopic linewidth of the pump pulse. Therefore, the Kerr-induced dynamic evolution is negligible within a time scale on the order of microseconds. At the more accurate Kerr-free point, the nonlinearity from the qubit ${K_{\rm qs}}/2\pi={\chi ^{\rm 2}}/4{K_{\rm q}}\approx-11\ {\rm kHz}$ can be compensated by $K_{\rm s}$, where the dynamic evolution is ideally eliminated. 

\begin{figure*}[h]
\centering
\includegraphics[width=0.9\textwidth]{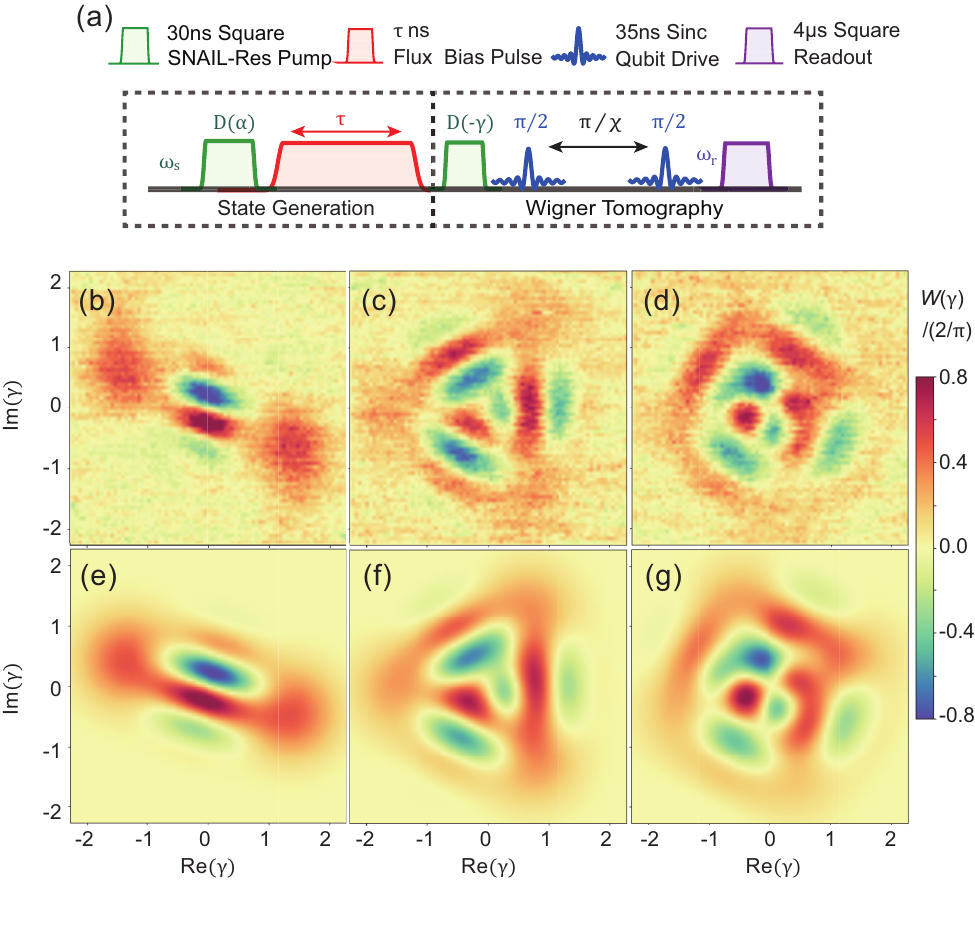}
\caption{Schrödinger cat states generation through the fast modulation of Kerr nonlinearity. (a) Pulse sequence for generating the $m$-component Poisson distributed cat states. (b)-(d) Measured Wigner functions of the $m$-component cat ($m$=2,3,4). (e-g) Numerical Wigner functions of the $m$-component cat ($m$=2,3,4) obtained through QuTip \cite{b38}.  }\label{Fig5}
\end{figure*}

In order to show the merits of preparing the quantum state at the Kerr-free point, as an example, we displace the nonlinear resonator with $\alpha {\rm{ = }}1.42$ at $\Phi_{\rm ext} = 0.4026\ {\Phi _{\rm 0}}\ (K\approx 0)$ and $\Phi_{\rm ext} = 0.41\ {\Phi _{\rm 0}}\ (K/2\pi\approx 0.5{\rm MHz})$, respectively. Then, we wait for a time duration $\Delta {t}$ before performing the Wigner tomography on the states by taking the parity measurements \cite{b29}\cite{b44}, where the pulse sequence is shown in Fig. \ref{Fig3}(a). The results [Fig. \ref{Fig3}(b-g)] clearly illustrate that the quantum states can be preserved well at Kerr-free point whereas the phase of the state collapses when the Kerr nonlinearity is nonzero. Moreover, the frequency shift among energy levels may induce variations of the photon distribution (as what we discussed in the two-tone nonlinearity measurement above). Injection of multiple photons would be much easier at the Kerr-free point because of the simple spectrum \cite{b30}. As a result, a larger Hilbert space of photons provides us with a larger coding area either for error correction \cite{b7}\cite{b14}\cite{b45} or loss suppression \cite{b17}. 


Furthermore, the fast tunablity of the Kerr coefficient can also be used to generate non-classical quantum states. The Kerr cat qubits \cite{b18} with the related error correction methods \cite{b31} benefiting from Kerr nonlinearity show the potential of dissipation-insensitive and long lifetime quantum computation in a multidimensional Hilbert space. 

For example, we consider the Bloch sphere of a Kerr-cat qubit which is constructed with a group of perpendicular states \cite{b18}:
\begin{equation}
\begin{aligned}
\lvert  {{\varphi_{\rm \pm X}}} \rangle&=\lvert  { \pm \alpha } \rangle,\\
\lvert  {{\varphi _{\rm \pm Z}}} \rangle&=\lvert  {  \alpha } \rangle\pm\lvert  {  -\alpha } \rangle,\\
\lvert  {{\varphi _{\rm \pm Y}}} \rangle&=\lvert  {  \alpha } \rangle\pm i\lvert  {  -\alpha } \rangle,\label{eq4}
\end{aligned}
\end{equation}

$\lvert  {{\varphi_{\rm \pm X}}} \rangle=\lvert  { \pm \alpha } \rangle$ are the coherent states generated by pumping our nonlinear resonator with coherent pulses. To prepare the cat states $\lvert  {{\varphi _{\rm \pm Y}}} \rangle=\lvert  {  \alpha } \rangle\pm i\lvert  {  -\alpha } \rangle$, the Kerr nonlinearity normally play an important role \cite{b18}\cite{b21}. As shown in Fig. \ref{Fig5}(a), the flux bias pulse (with a pulse width $\tau$) after the first displacement $D(\alpha )$ introduces a flux bias shift, as well as a large Kerr coefficient $K$. The coherent states evolution under this nonlinear Hamiltonian results in phase shifts among Fock states $\lvert  N\rangle$ ($N$=0,1,2,...), which means the rotating speed in the phase space is not uniform. Therefore, when we initialize the system with a coherent state $\lvert  {  \alpha } \rangle$, the evolution of the field states (during the flux bias pulse in Fig. \ref{Fig5}(a)) can be written as \cite{b18}:
\begin{equation}
\begin{aligned}
\lvert  {\Psi (\tau)}\rangle  &={e^{i\frac{K}{2}{{({a^\dag }a)}^{\rm 2}}\tau}}\lvert \alpha  \rangle \\
&= {e^{ - {{\lvert \alpha  \rvert }^{\rm 2}}/2}}\sum\limits_{ N}{\frac{{{\alpha ^N}}}{{\sqrt {N!} }}{e^{i\frac{K}{2}{N^{\rm 2}}\tau}}\lvert  N \rangle } \label{eq5}
\end{aligned}
\end{equation}

Under certain circumstances, the $m$-component cat states can be generated when the nonlinear evolution time $\tau  = {\tau _{\rm 0}}/m\  {\rm with}\ \tau _{\rm 0}=2\pi/K\ (m=1,2,...)$. In particular, when
\begin{equation}
\tau=\pi/K \quad {\rm or} \quad 3\pi /K ,\label{eqS17}
\end{equation}
the final state is 
\begin{equation}
\lvert  {\Psi (\tau)}\rangle=\lvert { \alpha  } \rangle  \pm i \lvert { -\alpha } \rangle.\label{eqS18}
\end{equation}

In our experiment, we chose $\alpha=1.42$ and a flux bias with $K/2\pi=5.21$ MHz corresponding to $\tau _0=192$ ns. When the pulse width of the flux bias is either $\tau  = {\tau _{\rm 0}}/2=96\ {\rm ns}$ or $\tau = 3{\tau _{\rm 0}}/2=288\ {\rm ns}$, we get $\lvert  {{\varphi _{\rm \pm Y}}} \rangle=\lvert  {  \alpha } \rangle\pm i\lvert  { -\alpha } \rangle$. By setting $\tau  = {\tau _{\rm 0}}/3\ (64\ {\rm ns})$ and $\tau  = {\tau _{\rm 0}}/4\ (48\ {\rm ns})$, we also implement 3- and 4- component cat states which can be used for quantum error correction \cite{b31}\cite{b37}. Wigner functions of 2, 3 and 4- component cat states are measured and shown in Fig. \ref{Fig5}(b-d).

The fidelity of the above Schrödinger's cat states can be calculated by comparing the measured Wigner function $W_{\rm meas}$ [Fig. \ref{Fig5}(b-d)] with the numerical one $W_{\rm cal}$ [Fig. \ref{Fig5}(e-g)]. The fidelity can be written as \cite{b36} 
\begin{equation}
F{\rm{ = }}\pi \int {{\rm{d}}{\gamma ^{\rm 2}}{W_{\rm meas}}(\gamma ){W_{\rm cal}}(\gamma )}, \label{eq6}
\end{equation}
where $\gamma$ is the displacement vector, integrated through the whole phase space.

Here, the fidelity $F_{m}$ of the $m$-component cat from our measurements is:
\begin{equation}
F_{\rm 2}=89.1\%,F_{\rm 3}=81.3\%,F_{\rm 4}=83.15\%. \label{eq7}
\end{equation}
Note that the distortion caused by the imperfect Wigner tomography has not been eliminated. Moreover, the fidelities are currently limited by our device coherent time $~1\ {\rm \mu s}$ (see Methods). To improve the performance of our device, we need to mention that the ancillary qubit is a nonnegligible source of the collapse and decoherence of the bosonic quantum states. It is therefore beneficial to reduce the impacts from the qubit (e.g. spectrally isolating the ancillary qubit while not in use). 

In previous strategies with a fixed Kerr coefficient \cite{b18}\cite{b21}, the $m$-component cat state is stabilized by applying a squeezing drive continuously. However, in our case, the cat states can be maintained in the Kerr-free system passively. Therefore, after the state generation, the system is immediately tuned back to the Kerr-free point, where our resonator can be described by a linear Hamiltonian \cite{b41}, leading to a better storage and evolution of the cat states within a desirable lifetime (see Methods). 

To verify the feasible controllability of our nonlinear resonator, we successfully generate an odd cat state, $\lvert  {{\varphi _{\rm - Z}}} \rangle=\lvert  {  \alpha } \rangle-\lvert  { -\alpha } \rangle$ by following the ancilla assisted cat preparation method \cite{b28}\cite{b33}. By using the spectral selectivity and different evolution induced by the dispersive shift, the odd cat state is obtained (Details see Methods). In summary, we have achieved the Y rotation $R_{\rm Y}$ and Z rotation $R_{\rm Z}$ and fulfilled the requirement of logic gates. It indicates that our platform can be regarded as a logical qubit.

In conclusion, we have generated nonclassical states through the fast tunable nonlinearity on a SNAIL-terminated resonator where the tunable range is up to 10 MHz. 
Compared to the cat states in 3D cavities \cite{b53} where the state preparation is based on the ancillary qubit, our method is more straightforward from the fast tuning of the Kerr coefficient of the nonlinear resonator itself. Thus, our scheme is much simpler and has no affect from the imperfect preparation on the ancillary qubit. Moreover, compared to the two-photon driving strategy \cite{b18}\cite{b21}\cite{b56}, the time to prepare the cat states is about 1/$K$ in our method, which is several times faster than the adiabatic case \cite{b21}. Meanwhile, at the Kerr-free point, the states of light can be stored passively without consecutive pump. Finally, our platform is more compact compared to 3D cavities, and shows the capability to integrate more elements. Therefore, our novel method shows a possibility of the extensible and low-crosstalk bosonic based quantum computation in the future.

Our method provides a new avenue to achieve continuous-variable quantum information processing. It can be used to achieve universal control of bosonic codes. One direct application of our circuitry is constructing hardware-efficient, loss-tolerable quantum computers with error correction codes \cite{b7}\cite{b53}. Furthermore, networks of coupled resonators can be used to achieve quantum annealing architectures \cite{b55} and quantum simulations such as phase transitions \cite{b47}, Gaussian boson sampling \cite{b54}, etc.

\bibliographystyle{unsrt}
\bibliography{reflist}

\begin{thebibliography}{10}

\bibitem{b1}
Frank Arute, Kunal Arya, Ryan Babbush, et~al.
\newblock Quantum supremacy using a programmable superconducting processor.
\newblock {\em Nature}, 574(7779):505--510, 2019.

\bibitem{b2}
Dieter Jaksch, Juan~Ignacio Cirac, Peter Zoller, et~al.
\newblock Fast quantum gates for neutral atoms.
\newblock {\em Phys. Rev. Lett.}, 85(10):2208, 2000.

\bibitem{b48}
Christopher Monroe and Jungsang Kim.
\newblock Scaling the ion trap quantum processor.
\newblock {\em Science}, 339(6124):1164--1169, 2013.

\bibitem{b49}
Colin~D Bruzewicz, John Chiaverini, Robert McConnell, et~al.
\newblock Trapped-ion quantum computing: {P}rogress and challenges.
\newblock {\em Applied Physics Reviews}, 6(2):021314, 2019.

\bibitem{b50}
Morten Kjaergaard, Mollie~E Schwartz, Jochen Braum{\"u}ller, et~al.
\newblock Superconducting qubits: {C}urrent state of play.
\newblock {\em Annual Review of Condensed Matter Physics}, 11:369--395, 2020.

\bibitem{b5}
Ming Gong, Shiyu Wang, Chen Zha, et~al.
\newblock Quantum walks on a programmable two-dimensional 62-qubit
  superconducting processor.
\newblock {\em Science}, 372(6545):948--952, 2021.

\bibitem{b51}
Austin~G. Fowler, Matteo Mariantoni, John~M. Martinis, et~al.
\newblock Surface codes: {T}owards practical large-scale quantum computation.
\newblock {\em Phys. Rev. A}, 86:032324, 2012.

\bibitem{b6}
Brian Vlastakis, Gerhard Kirchmair, Zaki Leghtas, et~al.
\newblock Deterministically encoding quantum information using 100-photon
  {S}chr{\"o}dinger cat states.
\newblock {\em Science}, 342(6158):607--610, 2013.

\bibitem{b7}
Philippe Campagne-Ibarcq, Alec Eickbusch, Steven Touzard, et~al.
\newblock Quantum error correction of a qubit encoded in grid states of an
  oscillator.
\newblock {\em Nature}, 584(7821):368--372, 2020.

\bibitem{b8}
Timo Hillmann, Fernando Quijandr{\'\i}a, G{\"o}ran Johansson, et~al.
\newblock Universal gate set for continuous-variable quantum computation with
  microwave circuits.
\newblock {\em Phys. Rev. Lett.}, 125(16):160501, 2020.

\bibitem{b42}
Ehud Altman, Kenneth~R Brown, Giuseppe Carleo, et~al.
\newblock Quantum simulators: {A}rchitectures and opportunities.
\newblock {\em PRX Quantum}, 2(1):017003, 2021.

\bibitem{b22}
Max Hofheinz, H~Wang, Markus Ansmann, et~al.
\newblock Synthesizing arbitrary quantum states in a superconducting resonator.
\newblock {\em Nature}, 459(7246):546--549, 2009.

\bibitem{b23}
Dietrich Leibfried, DM~Meekhof, BE~King, et~al.
\newblock Experimental determination of the motional quantum state of a trapped
  atom.
\newblock {\em Phys. Rev. Lett.}, 77(21):4281, 1996.

\bibitem{b57}
Christopher Chamberland, Kyungjoo Noh, Patricio Arrangoiz-Arriola, et~al.
\newblock Building a fault-tolerant quantum computer using concatenated cat
  codes.
\newblock {\em PRX Quantum}, 3(1):010329, 2022.

\bibitem{b11}
Yuwei Ma, Yuan Xu, Xianghao Mu, et~al.
\newblock Error-transparent operations on a logical qubit protected by quantum
  error correction.
\newblock {\em Nature Physics}, 16(8):827--831, 2020.

\bibitem{b12}
Daniel Gottesman, Alexei Kitaev, and John Preskill.
\newblock Encoding a qubit in an oscillator.
\newblock {\em Physical Review A}, 64(1):012310, 2001.

\bibitem{b13}
Wen-Long Ma, Shruti Puri, Robert~J Schoelkopf, et~al.
\newblock Quantum control of bosonic modes with superconducting circuits.
\newblock {\em Science Bulletin}, 66(17):1789--1805, 2021.

\bibitem{b52}
Seth Lloyd and Samuel~L Braunstein.
\newblock Quantum computation over continuous variables.
\newblock {\em Phys. Rev. Lett.}, 82(8):1784, 1999.

\bibitem{b15}
Gerhard Kirchmair, Brian Vlastakis, Zaki Leghtas, et~al.
\newblock Observation of quantum state collapse and revival due to the
  single-photon {K}err effect.
\newblock {\em Nature}, 495(7440):205--209, 2013.

\bibitem{b18}
Alexander Grimm, Nicholas~E Frattini, Shruti Puri, et~al.
\newblock Stabilization and operation of a {K}err-cat qubit.
\newblock {\em Nature}, 584(7820):205--209, 2020.

\bibitem{b16}
Zaki Leghtas, Steven Touzard, Ioan~M Pop, et~al.
\newblock Confining the state of light to a quantum manifold by engineered
  two-photon loss.
\newblock {\em Science}, 347(6224):853--857, 2015.

\bibitem{b20}
NE~Frattini, U~Vool, S~Shankar, A~Narla, KM~Sliwa, et~al.
\newblock 3-wave mixing {J}osephson dipole element.
\newblock {\em Applied Physics Letters}, 110(22):222603, 2017.

\bibitem{b27}
NE~Frattini, VV~Sivak, A~Lingenfelter, et~al.
\newblock Optimizing the nonlinearity and dissipation of a snail parametric
  amplifier for dynamic range.
\newblock {\em Physical Review Applied}, 10(5):054020, 2018.

\bibitem{b30}
Yong Lu, Marina Kudra, Timo Hillmann, et~al.
\newblock Resolving {F}ock states near the {K}err-free point of a
  superconducting resonator.
\newblock {\em arXiv preprint arXiv:2210.09718}, 2022.

\bibitem{b43}
Mark Dykman.
\newblock {\em Fluctuating nonlinear oscillators: from nanomechanics to quantum
  superconducting circuits}.
\newblock Oxford University Press, 2012.

\bibitem{b25}
Atsushi Noguchi, Alto Osada, Shumpei Masuda, et~al.
\newblock Fast parametric two-qubit gates with suppressed residual interaction
  using the second-order nonlinearity of a cubic transmon.
\newblock {\em Physical Review A}, 102(6):062408, 2020.

\bibitem{b39}
T.~Yamaji, S.~Kagami, A.~Yamaguchi, et~al.
\newblock Spectroscopic observation of the crossover from a classical {D}uffing
  oscillator to a {K}err parametric oscillator.
\newblock {\em Physical Review A}, 105, 02 2022.

\bibitem{b38}
R~Barends, J~Kelly, A~Megrant, et~al.
\newblock Superconducting quantum circuits at the surface code threshold for
  fault tolerance.
\newblock {\em Nature}, 508:500--3, 04 2014.

\bibitem{b29}
Samuel Deleglise, Igor Dotsenko, Clement Sayrin, et~al.
\newblock Reconstruction of non-classical cavity field states with snapshots of
  their decoherence.
\newblock {\em Nature}, 455(7212):510--514, 2008.

\bibitem{b44}
Uwe von L{\"u}pke, Yu~Yang, Marius Bild, et~al.
\newblock Parity measurement in the strong dispersive regime of circuit quantum
  acoustodynamics.
\newblock {\em Nature Physics}, pages 1--6, 2022.

\bibitem{b14}
Marina Kudra, Mikael Kervinen, Ingrid Strandberg, et~al.
\newblock Robust preparation of {W}igner-negative states with optimized
  {SNAP}-displacement sequences.
\newblock {\em PRX Quantum}, 3(3):030301, 2022.

\bibitem{b45}
William~P Livingston, Machiel~S Blok, Emmanuel Flurin, et~al.
\newblock Experimental demonstration of continuous quantum error correction.
\newblock {\em Nature communications}, 13(1):1--7, 2022.

\bibitem{b17}
C.~Berdou, A.~Murani, U.~R{\'{e} }glade, et~al.
\newblock One hundred second bit-flip time in a two-photon dissipative
  oscillator.
\newblock {\em {PRX} Quantum}, 4(2), jun 2023.

\bibitem{b31}
Nissim Ofek, Andrei Petrenko, Reinier Heeres, et~al.
\newblock Extending the lifetime of a quantum bit with error correction in
  superconducting circuits.
\newblock {\em Nature}, 536(7617):441--445, 2016.

\bibitem{b21}
Shruti Puri, Samuel Boutin, and Alexandre Blais.
\newblock Engineering the quantum states of light in a {K}err-nonlinear
  resonator by two-photon driving.
\newblock {\em npj Quantum Information}, 3(1):1--7, 2017.

\bibitem{b37}
Marcel Bergmann and Peter van Loock.
\newblock Quantum error correction against photon loss using multicomponent cat
  states.
\newblock {\em Physical Review A}, 94(4):042332, 2016.

\bibitem{b36}
AV~Chizhov, L~Kn{\"o}ll, and D-G Welsch.
\newblock Continuous-variable quantum teleportation through lossy channels.
\newblock {\em Physical Review A}, 65(2):022310, 2002.

\bibitem{b41}
Timo Hillmann and Fernando Quijandr{\'\i}a.
\newblock Designing {K}err interactions for {Q}uantum {I}nformation
  {P}rocessing via {C}ounterrotating {T}erms of {A}symmetric
  {J}osephson-{J}unction {L}oops.
\newblock {\em Physical Review Applied}, 17(6):064018, 2022.

\bibitem{b28}
Serge Haroche and J.~Raimond.
\newblock {\em Exploring the {Q}uantum: {A}toms, {C}avities, and {P}hotons}.
\newblock Oxford Univ Pr, 2006.

\bibitem{b33}
Craig~M Savage, Samuel~L Braunstein, and Daniel~F Walls.
\newblock Macroscopic quantum superpositions by means of single-atom
  dispersion.
\newblock {\em Optics Letters}, 15(11):628--630, 1990.

\bibitem{b53}
Barbara~M Terhal, Jonathan Conrad, and Christophe Vuillot.
\newblock Towards scalable bosonic quantum error correction.
\newblock {\em Quantum Science and Technology}, 5(4):043001, 2020.

\bibitem{b56}
Rapha{\"e}l Lescanne, Marius Villiers, Th{\'e}au Peronnin, et~al.
\newblock Exponential suppression of bit-flips in a qubit encoded in an
  oscillator.
\newblock {\em Nature Physics}, 16(5):509--513, 2020.

\bibitem{b55}
Wolfgang Lechner, Philipp Hauke, and Peter Zoller.
\newblock A quantum annealing architecture with all-to-all connectivity from
  local interactions.
\newblock {\em Science advances}, 1(9):e1500838, 2015.

\bibitem{b47}
MI~Dykman, Christoph Bruder, Niels L{\"o}rch, et~al.
\newblock Interaction-induced time-symmetry breaking in driven quantum
  oscillators.
\newblock {\em Physical Review B}, 98(19):195444, 2018.

\bibitem{b54}
Juan~M Arrazola, Ville Bergholm, Kamil Br{\'a}dler, et~al.
\newblock Quantum circuits with many photons on a programmable nanophotonic
  chip.
\newblock {\em Nature}, 591(7848):54--60, 2021.

\bibitem{b32}
Chen Wang, Yvonne~Y Gao, Philip Reinhold, et~al.
\newblock A {S}chr{\"o}dinger cat living in two boxes.
\newblock {\em Science}, 352(6289):1087--1091, 2016.

\bibitem{b40}
Hang Xue, Zhirong Lin, Jiang Wenbing, et~al.
\newblock Fabrication and characterization of all-{N}b lumped-element
  {J}osephson parametric amplifiers.
\newblock {\em Chinese Physics B}, 30, 03 2021.

\end{thebibliography}


\begin{thebibliography}{1}

\bibitem{b27}
N.~E. Frattini, V.~V. Sivak, A.~Lingenfelter, et~al.
\newblock Optimizing the nonlinearity and dissipation of a {SNAIL} parametric
  amplifier for dynamic range.
\newblock {\em Physical Review Applied}, 10(5), nov 2018.

\bibitem{b30}
Yong Lu, Marina Kudra, Timo Hillmann, et~al.
\newblock Resolving fock states near the kerr-free point of a superconducting
  resonator.
\newblock 10 2022.

\bibitem{b28}
Serge Haroche and J.~Raimond.
\newblock {\em Exploring the Quantum: Atoms, Cavities, and Photons}.
\newblock Oxford Univ Pr, 2006.

\bibitem{b32}
Chen Wang, Yvonne Gao, Philip Reinhold, et~al.
\newblock A schrodinger cat living in two boxes.
\newblock {\em Science}, 352, 01 2016.

\bibitem{b18}
A.~Grimm, N.~E. Frattini, S.~Puri, et~al.
\newblock Stabilization and operation of a kerr-cat qubit.
\newblock {\em Nature}, 584(7820):205--209, aug 2020.

\bibitem{b36}
A.~V. Chizhov, L.~Knöll, and D.-G. Welsch.
\newblock Continuous-variable quantum teleportation through lossy channels.
\newblock {\em Physical Review A}, 65(2), jan 2002.

\bibitem{b38}
J.R. Johansson, P.D. Nation, and Franco Nori.
\newblock Qutip 2: A python framework for the dynamics of open quantum systems.
\newblock {\em Computer Physics Communications}, 184(4):1234--1240, 2013.

\end{thebibliography}


\begin{thebibliography}{1}

\bibitem{b28}
Serge Haroche and J.~Raimond.
\newblock {\em Exploring the {Quantum: Atoms, Cavities, and Photons}}.
\newblock Oxford Univ Pr, 2006.

\bibitem{b32}
Chen Wang, Yvonne Gao, Philip Reinhold, et~al.
\newblock A {Schrodinger}{ Cat Living in Two Boxes}.
\newblock {\em Science}, 352, 01 2016.

\bibitem{b40}
Hang Xue, Zhirong Lin, Jiang Wenbing, et~al.
\newblock Fabrication and characterization of all-{N}b lumped-element josephson
  parametric amplifiers.
\newblock {\em Chinese Physics B}, 30, 03 2021.

\end{thebibliography}

\begin{figure*}[!h]
	\centering
	\includegraphics[width=0.9\textwidth]{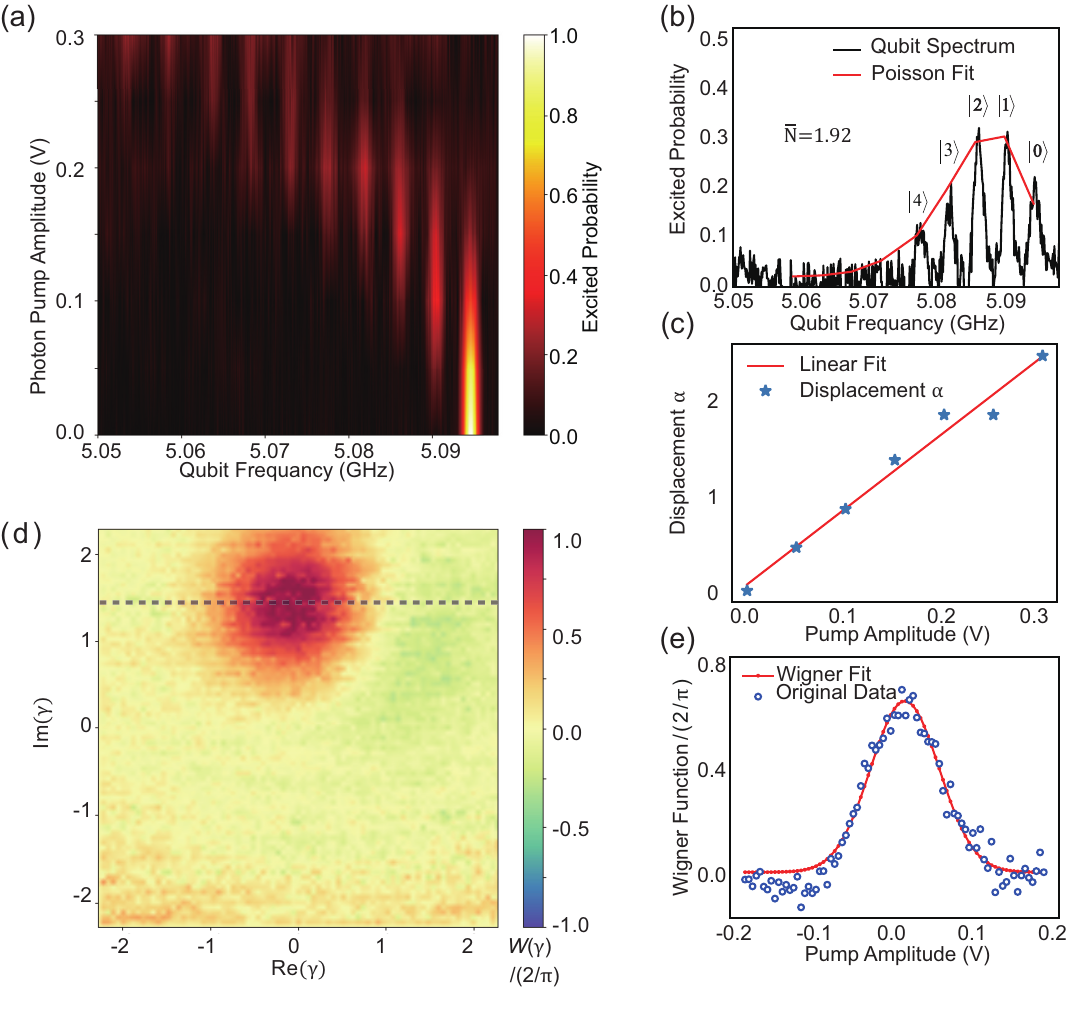}
	\caption{Two strategies for photon number calibration (a) Qubit spectroscopy under different pump pulse amplitudes. (b) Poisson fitting with an average photon number 1.92. (c) Linear relationship between displacement and pump amplitude. (d) Wigner function of a coherent state with $\alpha = 1.42$. (e) Fitting of Wigner function (dashed grey line in (d)).}\label{FigS1}

\end{figure*}

\begin{figure}[!ht]
\centering
\includegraphics[width=0.45\textwidth]{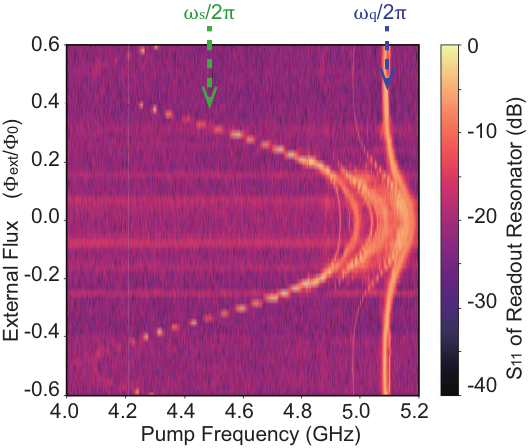}
\caption{Flux modulated frequency of the SNAIL-terminated resonator and ancillary qubit, obtained by the reflection coefficient measurement on the readout resonator with different pump frequency.}\label{FigS2}
\end{figure}

\begin{figure*}[!ht]
\centering
\includegraphics[width=0.9\textwidth]{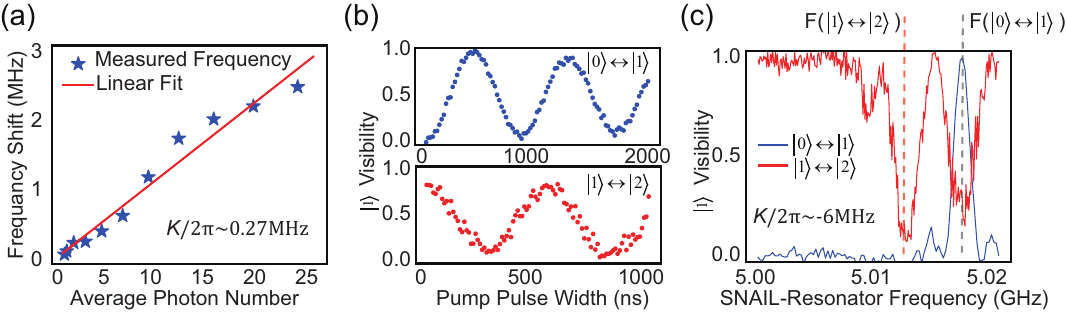}
\caption{Experimental results of 1-tone and 2-tone nonlinearity measurement. (a) Relationship between the frequency shift and the photon number in 1-tone measurement with external flux $\Phi_{\rm ext} \sim0.409{\Phi _{\rm 0}}$.  (b) Rabi oscillation between the Fock states ($\lvert 0\rangle  \leftrightarrow \lvert 1 \rangle $and $\lvert 1\rangle  \leftrightarrow \lvert 2 \rangle $). (c) Frequency scan of the SNAIL-terminated resonator under a 2-tone measurement with a conditional $\pi$ pulse for $\lvert 1\rangle$\ ($\Phi_{\rm ext}  \sim0.24{\Phi _{\rm 0}}$).}\label{FigS3}
\end{figure*}

\section*{Acknowledgements}\label{sec3}
The authors acknowledge the use of the Nanofabrication Laboratory (NFL) at Chalmers.
We wish to express our gratitude to Xiaoming Xie, Lars J\"{o}nsson, Fernando Quijandr{\'\i}a, Timo Hillmann, and Hang-Xi Li for help. 
This work is supported in part by the Shanghai Technology Innovation Action Plan Integrated Circuit Technology Support Program (No. 22DZ1100200), the National Natural Science Foundation of China (No. 92065116), Strategic Priority Research Program of the Chinese Academy of Sciences (Grant No. XDA18000000), and the Key-Area Research and Development Program of Guangdong Province, China (No. 2020B0303030002). Y.L. and P.D. acknowledge support from the Knut and Alice Wallenberg Foundation via the Wallenberg Center for Quantum Technology (WACQT) and from the Swedish Research Council (Grant number 2015-00152).

\section*{Author contributions}\label{sec4}
X.L.H. and Z.R.L. conceived the experiment. X.L.H. performed the experiments and analyzed the data. Y.L. triggered the project, designed, simulated, fabricated the device and helped with the measurement and analysis. D.Q.B. provided the theoretical support. X.L.H. wrote the manuscript together with Y.L. and Z.R.L. All authors contributed to the discussion of the results and proofreading of the manuscript. Z.R.L. supervised the project.

\section*{Methods}\label{sec2}

\subsection*{Photon number calibration}\label{subsec1}

The photon number in the resonator is measured by the spectroscopy of the ancillary qubit. Here, we pump the resonator with a coherent pulse $D(\alpha)$. As shown in Fig. \ref{FigS1}(a), due to the dispersive coupling to the qubit, the qubit frequency has a Poisson distribution related to the photon pump amplitude $V$, where the $ N^{\rm{th}}$ peak (away from the native qubit frequency) corresponds to the probability of the Fock state $\lvert N \rangle$ with photon number $N$. By fitting the multi-peak spectra to a Poisson-distribution function, we can extract the average photon number $\overline N = {\alpha ^{\rm2}}$. Therefore, we can figure out the linear relationship between the photon pump amplitude $V$ and the values of $\alpha$, namely, $\alpha  = G  \cdot V$, where $G$ is the scale factor between displacement $\alpha$ and amplitude $V$.

In addition, to calibrate the photon more precisely under a low photon number ($\overline N<5$), we can also get the relationship $\alpha  = G \cdot V$ by fitting the Wigner distribution of the coherent state with [Fig. \ref{FigS1} (e)]: 
\begin{equation}
\begin{aligned}
W &= \frac{2}{\pi }\exp ( - 2{(\alpha  - {\alpha _0})^2}) \\
    &= \frac{2}{\pi }\exp ( - 2{(G \cdot V - {\alpha _0})^2}),\label{eqS22}
\end{aligned}
\end{equation}
where $\alpha_0$ is the initial displacement. 

These relationships are used for the nonlinearity, Wigner function and lifetime measurements discussed in the main text. However, because of the flux-dependent nonlinearity, the photon-number distribution does not satisfy Poisson function when the frequency shift $N \cdot K$ is comparable with the linewidth of the pump pulse ($\approx 2\ {\rm MHz}$ for the $500\ {\rm ns}$ pump pulse). The photon number calibration methods above is therefore only available for $\Phi  \sim0.4{\Phi _{\rm 0}}$ with $\lvert K/2\pi \rvert <2\ {\rm MHz}$.

\begin{figure*}[h]
\centering
\includegraphics[width=0.9\textwidth]{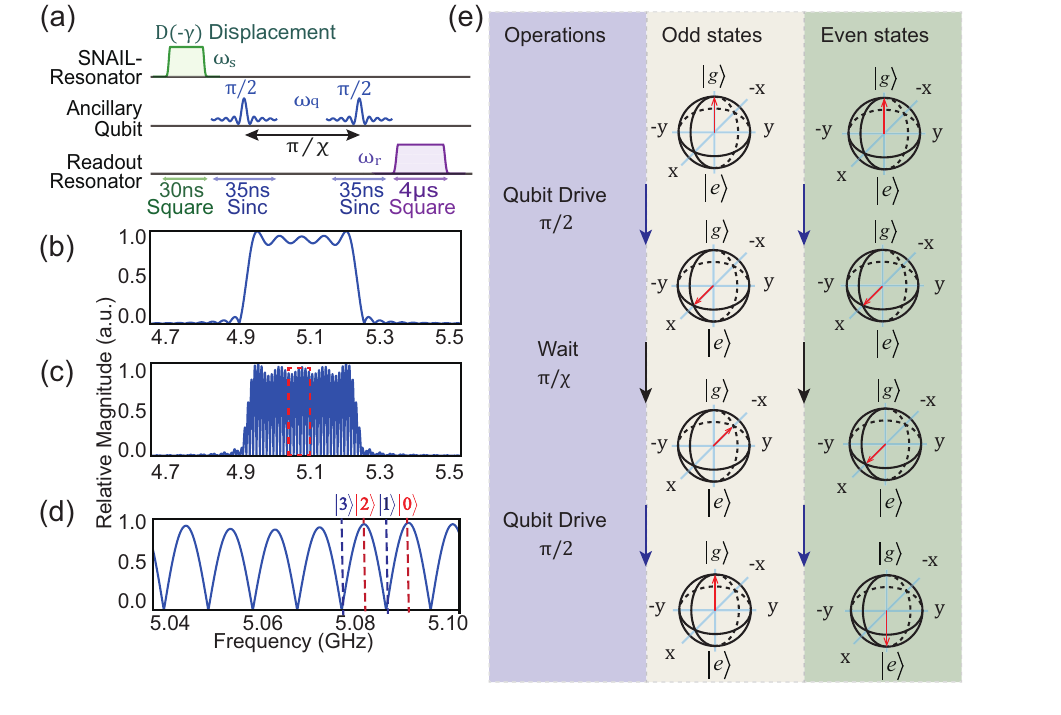}%
\caption{Wigner tomography. (a) Pulse sequence for Wigner tomography. (b) Frequency spectrum of a sinc shaped pulse. (c) Frequency spectrum of two sinc pulses with a time spacing $\pi/\chi$. (d) Enlarged view of (c). (e) Bloch sphere of the ancillary qubit during the cat state Wigner tomography.}\label{FigS5}
\end{figure*}

\subsection*{Nonlinearity characterization}\label{subsec2}

The Hamiltonian of the SNAIL element with three large Josephson junctions (with Josephson energy $E_{\rm J}$) and one smaller junction ($\beta E_{\rm J}$) can be written as (same as Eqn. \ref{eq1}):
\begin{equation}
\begin{split}
{U_{\rm SNAIL}}(\varphi_{\rm s} )=- \beta {E_{\rm J}}\cos (\varphi_{\rm s} )-3{E_{\rm J}}\cos (\frac{{{\varphi _{\rm ext}}- \varphi_{\rm s} }}{3}),\label{eqS1}
\end{split}
\end{equation}
where $\varphi _{\rm ext}$ is the external magnetic flux induced phase through the SNAIL loop,  $\varphi _{\rm s}$ is the phase different between the two ports of SNAIL. Coupled with a resonator (represented by capacity C and inductance L):

\begin{equation}
\begin{aligned}
H &= C\frac{{{\Phi _{\rm 0}}^{\rm2}}}{2}{{\dot \varphi }^{\rm 2}} + U(\varphi ,{\varphi _{\rm s}}),\\
U(\varphi ,{\varphi _{\rm s}}) &= \frac{1}{2}{E_{\rm L}}{(\varphi  - {\varphi _{\rm s}})^{\rm 2}} + {U_{\rm SNAIL}}({\varphi _{\rm s}}),\label{eqS2}
\end{aligned}
\end{equation}
where $\varphi$ is the mode canonical phase coordinate and ${E_{\rm L}} = {\Phi _{\rm 0}}^{\rm2}/L$ is the inductive energy. After Taylor expansion around the minimum point of potential $U$, the Hamiltonian of second quantization is \cite{b27}\cite{b30}:
\begin{equation}
{H_{\rm SNAIL-Res}} = \hbar {\omega _{\rm s}}a^\dag a + {g_{\rm 3}}{(a + a^\dag )^{\rm 3}} + {g_{\rm 4}}{(a + a^\dag )^{\rm 4}},\label{eqS3}
\end{equation}
where
\begin{equation}
\begin{aligned}
\hbar \omega_{\rm s} &= \sqrt {8{E_{\rm C}}{E_{\rm J}}{c_{\rm 2}}}, \\
\hbar {g_{\rm 3}} &= {c_{\rm 3}}\sqrt {{E_{\rm C}}\hbar {\omega _{\rm s}}} /6{c_{\rm 2}},\\
\hbar {g_{\rm 4}} &= {c_{\rm 4}}Ec/12{c_{\rm 2}},\\
{c_{\rm j}} &= \frac{1}{{{E_{\rm J}}}}\frac{{{d^{\rm j}}U}}{{d{\varphi ^{\rm j}}}}\lvert {\begin{array}{*{20}{c}}
{}\\
{{\varphi _{\rm m}}}
\end{array}} ,\\
Ec &= {e^{\rm 2}}/2C.\label{eqS4}
\end{aligned}
\end{equation}
The energy level n is:
\begin{equation}
{E_{\rm n}}/\hbar  = n{\omega _{\rm s}} + 6({g_{\rm 4}} - 5{g_{\rm 3}}^{\rm 2}/{\omega _{\rm s}})n(n + 1).\label{eqS5}
\end{equation}
Thus, the Kerr coefficient of SNAIL-terminated resonator can be written as:
\begin{equation}
\hbar K = \frac{{{d^{\rm 2}}{E_{\rm n}}}}{{d{n^{\rm 2}}}} = 12\hbar ({g_{\rm 4}} - \frac{{5{g_{\rm 3}}^{\rm 2}}}{{\omega_{\rm s}}}).\label{eqS6}
\end{equation}
The design parameters $\beta$ and $E_{\rm J}$, also the relationship between injected current and flux $\varphi_{\rm ext}$ can be extracted by fitting the flux modulation of the SNAIL-terminated resonator frequency(Fig. \ref{FigS2}) with Eqn. \ref{eqS5} \cite{b30}.

We developed two methods for the Kerr coefficient characterization, namely, single-tone and two-tone measurements. The single-tone measurement is only suitable for smaller Kerr coefficient (e.g. $\lvert  {K/2\pi } \rvert < 2\ {\rm MHz}$). To contrast, larger $K$ brings more frequency shift with a similar photon number. When the frequency shift is larger than the linewidth of the single pulse, the nonlinearity can be measured by the two-tone method.

\begin{figure*}[h]
\centering
\includegraphics[width=0.9\textwidth]{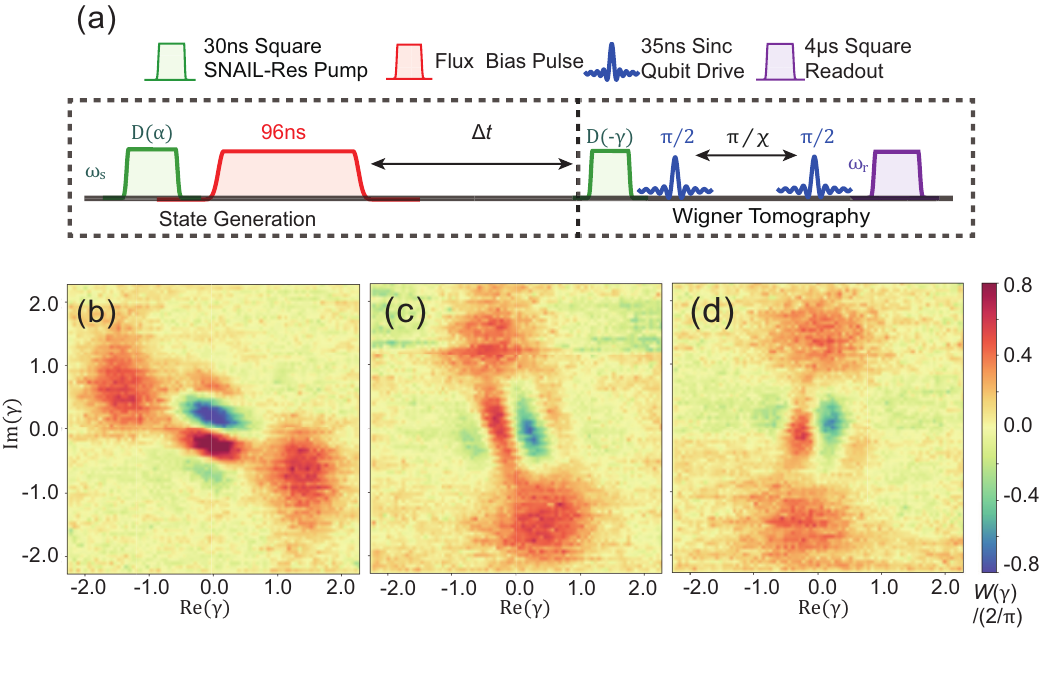}
\caption{Cat states generation and preservation.  (a) Pulse sequence for 2-component cat $\lvert { \alpha  } \rangle  \pm i \lvert { -\alpha } \rangle$ preparation. After preparation, the states are measured after $\Delta t=$ (b) 0 ns, (c)100 ns, (d)200 ns.}\label{FigS6}
\end{figure*}

In a single-tone measurement, in order to keep the frequency sensitivity, we choose a relatively long pulse with a pulse length up to 500 ns, then we sweep the pulse frequency with following a conditional $\pi$-pulse which can excite the qubit only if the cavity is empty. Therefore, it can be regarded as a photon probe. With different pump powers, we can see the shift of resonant frequency $\Delta f$ which obeys: 
\begin{equation}
\begin{aligned}
K/2\pi  &= {f_{\rm N}} - {f_{\rm N - 1}},\\
K/2\pi & = \Delta f/\overline N,\label{eqS7}
\end{aligned}
\end{equation}
where $\overline N $ is the average photon number. For example, in Fig. \ref{Fig2}(c), we see the frequency shift with different displacement (i.e. pump amplitude). By linearly fitting the relationship of the frequency shift and the average photon number, we get the Kerr coefficient[Fig. \ref{FigS3}(a)].
If $K$ is too large ($\lvert  {K/2\pi } \rvert >2\ {\rm MHz}$), however, it is not possible to cover the frequency range with a single tone. Then, we treat the SNAIL-terminated resonator as a three level system with an anharmonicity $K$. To verify it, Rabi oscillations between energy levels ($\lvert 0\rangle  \leftrightarrow \lvert 1 \rangle $and $\lvert 1\rangle  \leftrightarrow \lvert 2 \rangle $) are measured [Fig. \ref{FigS3}(b)]. Here, the conditional qubit $\pi$ pulse is also modified to be only valid for $\lvert 0 \rangle $,$\lvert 1 \rangle $ or $\lvert 2 \rangle $, respectively. Thus, $K$ is represented by the frequency difference of the first two pulses as
\begin{equation}
K/2\pi = f(\lvert 1 \rangle  \leftrightarrow \lvert 2\rangle ) - f(\lvert0\rangle  \leftrightarrow \lvert 1 \rangle ).\label{eqS8}
\end{equation}
The measurement results from single-tone and two-tone methods agree with the theoretical calculation very well [Fig. \ref{Fig2}(a)].

\subsection*{Wigner tomography}\label{subsec4}
The Wigner function of the states is obtained by the parity measurement. For definition, Wigner function can be described as:
\begin{equation}
\begin{aligned}
W(\gamma ) &= \frac{2}{\pi }Tr[D( - \gamma )\rho D(\gamma )P],\\
P &= {e^{i\pi {a^\dag }a}} = {( - 1)^{\rm N}},\label{eqS9}
\end{aligned}
\end{equation}

By applying a displacement $D(-\gamma)$ to a density operator $\rho$, we get a new state with density operator:
\begin{equation}
\rho ' = D( - \gamma )\rho D(\gamma ).\label{eqS10}
\end{equation}

Thus, the Wigner function $W(\gamma)$  is proportional to the average of parity operator $P$ which can be measured with the sequence in Fig. \ref{FigS5}(a). Here, we apply two $\pi/2$ pulses to the ancillary qubit. With a time spacing $\pi/\chi$ between two pules, this sequence can be treated as a parity measurement where the state of qubit$\lvert g\rangle$ ($\lvert e\rangle$) corresponds to $P=-1(1)$. The mechanism can be described in the qubit Bloch sphere [Fig. \ref{FigS5}(e)] and signal spectrum [Fig. \ref{FigS5}(b-d)]. The sequence to qubit is  $\pi$ (2$\pi$ ) pulse if photon number $N$ is even (odd). Considering the frequency shift of qubit, we employ a function of ${\rm sinc}(t) = \sin (t)/t$ to cover a larger spectrum range uniformly.


\subsection*{States preservation}\label{subsec7}
After the process of states generation, the cat states are preserved by turning the flux bias off. Here, the Wigner functions are measured after a certain evolution time $\Delta t$. As shown in Fig. \ref{FigS6}(b-d), the Kerr induced evolution is ideally prevented, the fidelity of 2-component cat ($\lvert { \alpha  } \rangle  \pm i \lvert { -\alpha } \rangle$) state is 89.1\%, 81.9\% and 75.8\% after (0, 100, 200ns).

\subsection*{Odd/Even Cat states generation}\label{subsec5}
\begin{figure}[h]
\centering
\includegraphics[width=0.45\textwidth]{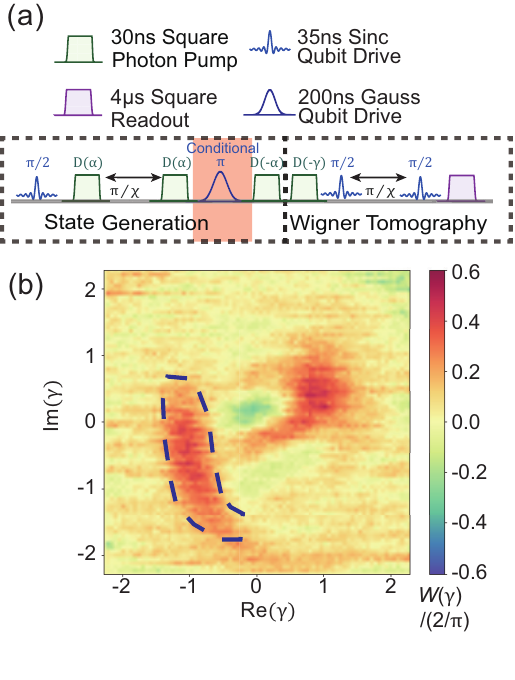}
\caption{Odd/even cat generation. (a) Pulse sequence for odd/even cat preparation. (b) Measured Wigner function of an odd cat state  $\lvert  {{\varphi _{\rm - Z}}} \rangle=\lvert  {  \alpha } \rangle-\lvert  {  -\alpha } \rangle$, where $\alpha {\rm{ = }}1.42$. }\label{Fig4}
\end{figure}
\begin{figure}[h]
\centering
\includegraphics[width=0.45\textwidth]{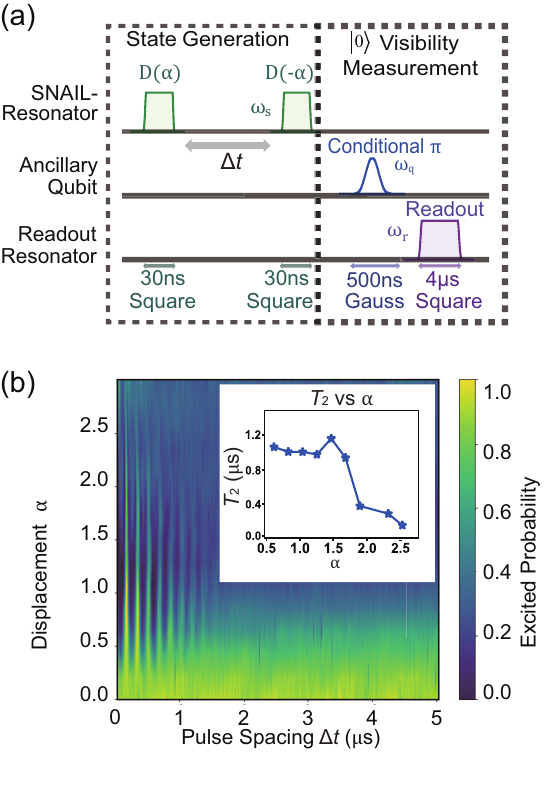}
\caption{Decoherent time $T_{\rm 2}$ measurement over different displacement $\alpha$. (a) Measurement pulse sequence. (b) The displacement (photon number) related $T_{\rm 2}$ measurement.}\label{FigS7}
\end{figure}

\begin{figure*}
\centering
\includegraphics[width=0.8\textwidth]{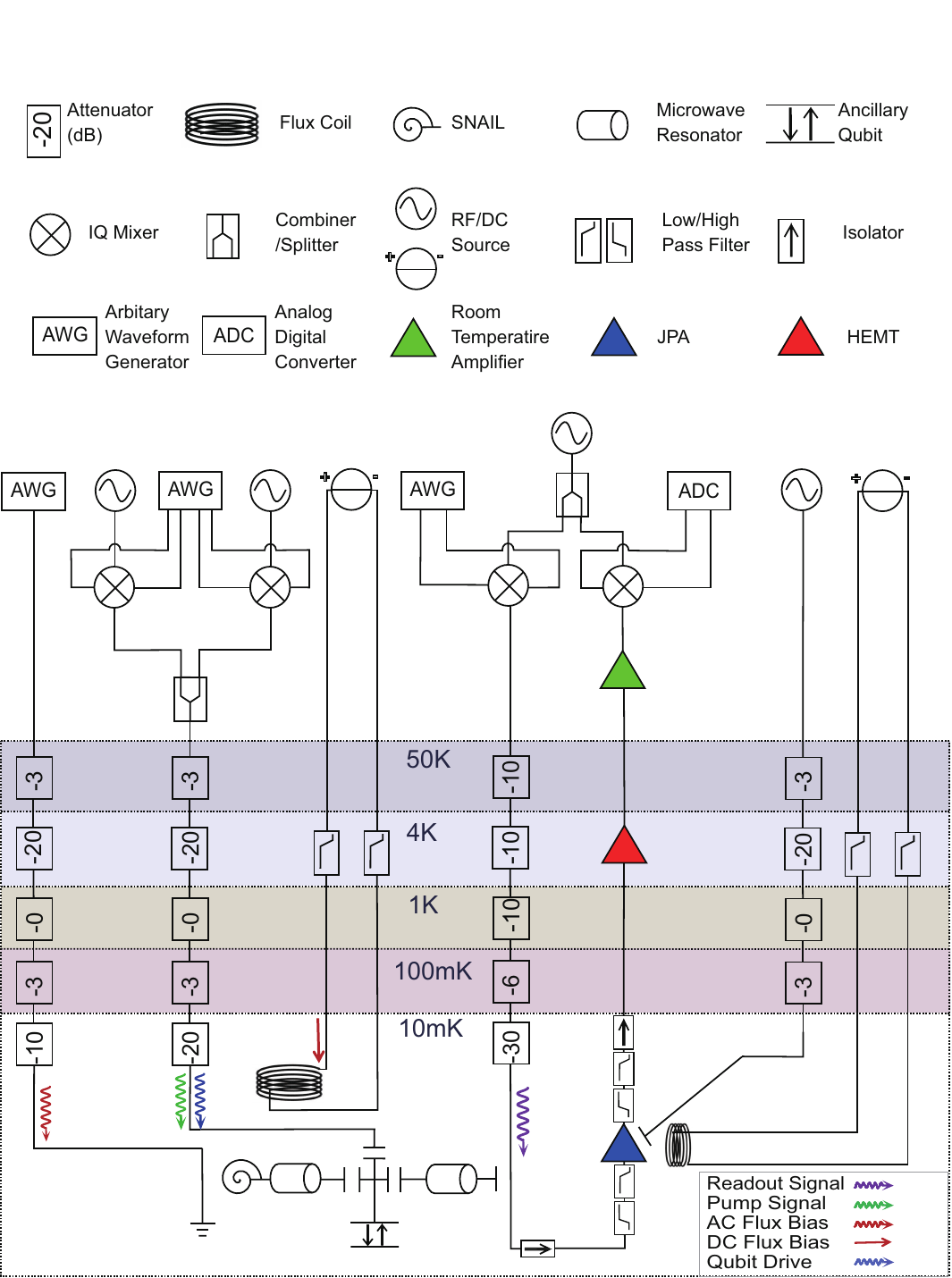}
\centering
\caption{\centering{Schematic diagram of the measurement system.}}\label{FigS4}

\end{figure*}

The cat states with odd/even properties are generated by following the pulse sequence in Fig. \ref{Fig4}(a) \cite{b28}\cite{b32}. We firstly excite the qubit to $(\lvert e\rangle  + \lvert g \rangle)/2 $ by a $\pi/2$-pulse. Then, a displacement pulse is injected to initialize the SNAIL-terminated resonator in a coherent state. After a certain period of evolution  $\tau\ (\tau=\pi/\chi,\ 3\pi/\chi)$. The entangled states can be written as:

\begin{equation}
\begin{aligned}
\lvert \Psi  \rangle  &= \frac{1}{{\sqrt 2 }}\lvert e \rangle  \otimes \lvert { - \alpha } \rangle  - \frac{1}{{\sqrt 2 }}\lvert g \rangle  \otimes \lvert \alpha  \rangle ,\ (\tau  = \pi /\chi )\\
\lvert \Psi  \rangle  &= \frac{1}{{\sqrt 2 }}\lvert e \rangle  \otimes \lvert { - \alpha } \rangle  +\frac{1}{{\sqrt 2 }}\lvert g \rangle  \otimes \lvert \alpha  \rangle ,\ (\tau  = 3\pi /\chi) \label{eqS11}
\end{aligned}
\end{equation}
After an additional displacement pulse $D(\alpha)$, the state becomes:
\begin{equation}
\lvert \Psi  \rangle  = \frac{1}{{\sqrt 2 }}\lvert e \rangle  \otimes \lvert { 0 } \rangle  \pm \frac{1}{{\sqrt 2 }}\lvert g \rangle  \otimes \lvert 2\alpha  \rangle. \label{eqS12}
\end{equation}
Next, we apply a conditional $\pi$-pulse, which only works when the photon number is small ($ < 4\alpha^2 $).
\begin{equation}
\lvert \Psi  \rangle  =\lvert g \rangle  \otimes (\lvert { 0 } \rangle  \pm\lvert { 2\alpha } \rangle ).\label{eqS13}
\end{equation}
Substantially, the two component cat states $\lvert { \alpha  } \rangle  \pm\lvert { -\alpha } \rangle$ are generated after the third displacement $D(-\alpha)$.
\begin{equation}
\lvert \Psi  \rangle  = \lvert g \rangle  \otimes (\lvert {  \alpha } \rangle  \pm\lvert { -\alpha } \rangle ).\label{eqS14}
\end{equation}

The corresponding Wigner function in Fig. \ref{Fig4}(b) clearly show the interference among coherent states. The asymmetric pattern, highlighted by the dashed blue line in Fig. \ref{Fig4}(b), comes from the higher decoherence and dissipation rates under a larger photon number (Fig. \ref{FigS7}). In details, during a period of the state preparation (shadowed by red in Fig. \ref{Fig4}(a)), a part of the state is displaced to the position $\lvert  { 2 \alpha } \rangle$, with the photon number four time larger than the initial state $\lvert  {  \alpha } \rangle$, leading to an enhanced decoherence and dissipation \cite{b28}. As a proof, the photon number dependent decoherence rate is measured (discussed in the next section).




\subsection*{Photon number related decoherence rate}\label{subsec6}
As theoretical analysis \cite{b28}, the decoherence and relaxation rate are highly affected by the photon number. The decoherent time of the light field is characterized by scanning the time spacing between two opposite displacement ($D(\alpha)$ and $D(-\alpha)$). In our system, we can clearly see the decoherent time $T_{\rm 2}$ drops while increasing the pump power (Fig. \ref{FigS7}). It is also the reason for the asymmetric shape in Fig. \ref{Fig4}(b).

\subsection*{Measurement setup}\label{subsec3}
The sample is mounted inside a dilution refrigerator (BlueFors LD400) system, the schematic diagram is shown in Fig. \ref{FigS4}. We use 3 layers of $\mu$-metal to shield the magnetic noise. A superconducting coil installed upon the sample holder to produce the static magnetic field. To amplify the output signal, we use three-level amplification-- a Josephson parametric amplifier (JPA) with Nb trilayer Josephson junctions \cite{b40} at 10 mK, a high electron mobility transistor (HEMT) at 4 K and a low-noise amplifier at room temperature.

\end{document}